\documentclass[3p,times,procedia]{elsarticle}
\usepackage{nupha_ecrc}

\volume{00}

\firstpage{1}

\journalname{Nuclear Physics A}

\runauth{A. Kurkela}

\jid{nupha}

\jnltitlelogo{Nuclear Physics A}

\usepackage{amssymb}

\usepackage[figuresright]{rotating}

\begin{document}

\begin{frontmatter}


\title{Isotropization and thermalization in heavy-ion collisions\tnoteref{label1}}

\dochead{}

\title{Initial state of Heavy-Ion Collisions: Isotropization and thermalization}


\author[cern,stav]{Aleksi Kurkela}

\address[cern]{Theoretical Physics Department, CERN, Geneva, Switzerland}
\address[stav]{Faculty of Science and Technology, University of Stavanger, Norway}

\begin{abstract}
I discuss how local thermal equilibrium and hydrodynamical flow are reached in heavy-ion collisions
in the weak coupling limit. 
\end{abstract}


\end{frontmatter}


\section{Introduction}

Over the past decade we have seen a remarkable phenomenological success 
of describing heavy-ion collisions using relativistic hydrodynamics. In the modern
theoretical view, hydrodynamics is seen as a low energy effective 
theory of the underlying quantum-field theory, and as such it is relying on an assumption
that the energy-momentum tensor can expanded around its thermal equilibrium value
$T_{eq.}$ in terms of gradients of flow fields,
\begin{equation}
T^{\mu\nu} \approx T_{eq.}^{\mu\nu}- 2 \eta  \nabla^{<\mu}u^{\nu>} +\ldots,
\end{equation}
where the terms in the series are graded by the number of derivatives
acting on the flow fields $u(x)$ \cite{Baier:2007ix}. If the flow fields are sufficiently
smooth, the flow is well described in terms of the few first terms in the series,
but when the gradients are large, the system is no longer described in terms
of hydrodynamics. 

Due to the singular geometry of an ultra-relativistic heavy-ion collision, the gradients of the 
flow fields diverge when approaching $\tau \rightarrow 0^+$,  the moment when the two Lorentz
contracted nuclei first collided. Therefore it is clear that the full time evolution of the 
collision is not amendable for hydrodynamical description and any hydrodynamical
simulation has to start at some \emph{initialization time}  $\tau_i$ after the collision, 
when the hydrodynamical description has become appropriate. The dynamics before
this time are \emph{prethermal} dynamics. 

The minimal requirement on the prethermal model is that it smoothly and automatically
approaches hydrodynamical flow. If this is the case, then the physical results become
independent of the unphysical choice of initialization time. This, however, is not the 
case in the vast majority of current implementations where the prethermal evolution is
either completely neglected, 
or modelled in a way that does not contain the
correct physics to reach hydrodynamical flow. 
The residual dependence
promotes the unphysical initialization time to a physical model parameter often referred as the \emph{thermalization time} which needs to be 
fixed by comparison to data. 
The problem is not only phenomenological. From the theoretical point of view, the failure to
reach correct behaviour is a sign of failure to identify the correct dominant physics.
\begin{figure}
\begin{center}
\includegraphics[width=0.4\textwidth]{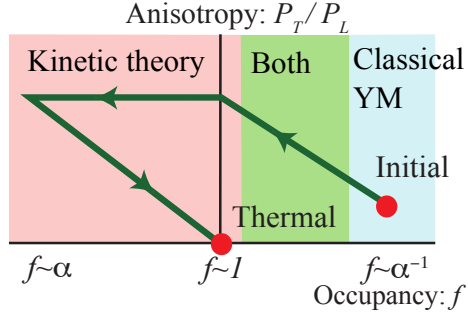}
\end{center}
\caption{
\label{fig:route}
A cartoon of the path from Color-Glass-Condensate initial condition to the thermal equilibrium in 
a plane of occupancy and anisotropy. At proper time $\tau \sim Q_s^{-1}$, the system is parametrically
overoccupied. Initially, the interactions are not strong enough to counter the expansion, and the system
becomes more anisotropic and dilute. Once the modes become underoccupied the dominant physics
change qualitatively and the system does not become anymore anisotropic. Eventually the underoccupied
system reaches equilibrium through the process of radiative break-up that causes the trajectory eventually
turn towards the origin in the figure. The system is described by classical field theory in the overoccupied region,
and by effective kinetic theory both in the underoccupied and overoccupied side, except when occupancies are
non-perturbatively large $f\sim 1/\alpha$.
}
\end{figure}

There have been two quite different strategies to attempt to quantitatively understand the 
prethermal evolution. On one hand, with
holographic methods, it has become possible to follow the far-from-equilibrium
evolution of certain gauge theories ($\mathcal{N}=4$ SYM in the $N_c\rightarrow \infty$ limit)
and observe the onset of hydrodynamical flow in collisions of sheets or lumps of supersymmetric matter \cite{Chesler:2008hg, Chesler:2010bi}. This has made it possible to 
construct a model that explicitly exhibits insensitivity to 
initialization time \cite{vanderSchee:2013pia}.

 The holographic method is, however, critically limited to a very special set of theories and cannot 
be applied to QCD. The complementary strategy has been to study
the prethermal evolution in QCD in the weak coupling limit (corresponding to the limit
of $\sqrt{s}\rightarrow \infty$), where in principle we expect to have analytical control of the theory.
In this limit, the initial condition for the prethermal evolution is fairly well understood in
terms of the physics of saturation and the Color-Glass-Condensate framework \cite{Gelis:2010nm}.
There have been several weak coupling motivated or inspired models used for phenomenology,
but none of these have so far quantitatively included all the necessary physics for a 
leading order calculation. However, in the recent past there has been significant progress, that will
 be reviewed in these proceedings. 

At weak coupling the smallness of coupling provides us with large scale separations. 
While the task of following the time evolution of a full quantum field theory is an unsolved
problem, these scale separations make it possible to study the prethermal evolution within different
effective theories, in particular using classical field theory and effective kinetic theory.
But while the scale separations make it possible to follow the time evolution of the system, they also 
make the prethermal evolution go through several stages of very different physics, all of which
must be quantitatively understood for a full leading order description of the prethermal evolution.

This proceeding is organized as follows. In Sec.~\ref{sec:route} I describe the overall features of the  thermalization process
and what are the roles of classical field theory and effective kinetic theory in its description. In the following Sections \ref{sec:clas} 
and  \ref{sec:kin} I discuss briefly recent simulations in classical Yang-Mills theory and in effective kinetic theory. 
I conclude with a discussion about the relevance of weak coupling thermalization to future phenomenology.

\section{Route to equilibrium, degrees of freedom, and the weak coupling strategy}
\label{sec:route}
\begin{figure}
\begin{center}
\includegraphics[width=0.5\textwidth]{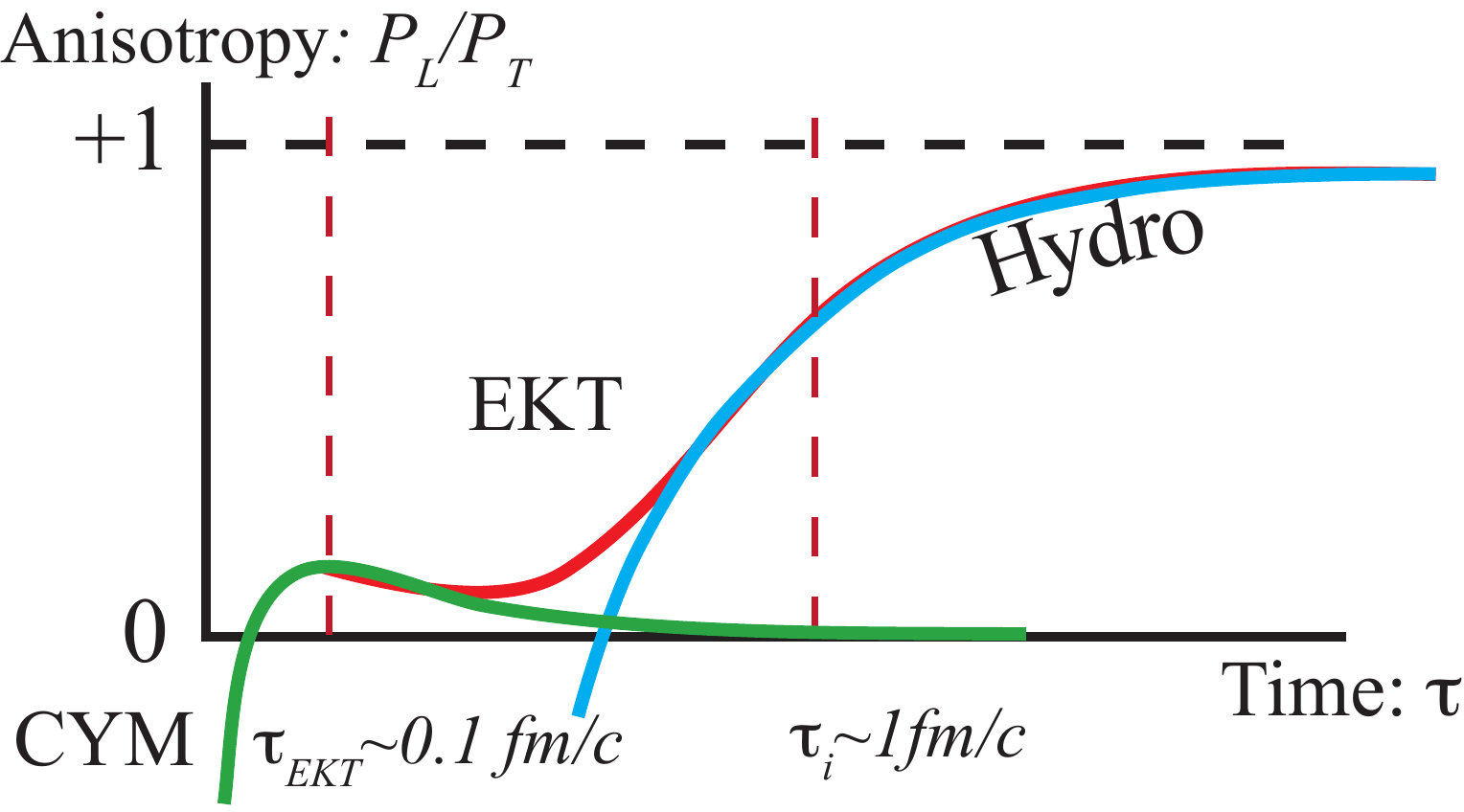}
\end{center}
\caption{
\label{fig2}
A cartoon of the strategy to bring the highly overoccupied Color-Glass-Condition initial condition to hydrodynamical region by a combination 
of classical Yang-Mills (CYM) simulations and effective kinetic theory (EKT). At early times the occupancies are non-perturbative $f\sim 1/\alpha$ 
the system is described in classical YM theory. The CYM evolution however makes the system ever more dilute and anisotropic never approaching 
local thermal equilibrium and hydrodynamic description. When the occupancies are $1\gg f \gg 1/\alpha_s$ both CYM and EKT give equally good leading
order description and the system can be smoothly passed from classical fields to particles at arbitrary time $\tau_{EKT}$. The subsequent evolution
in EKT will then deviate from the classical field evolution once the occupancies become close to unity $f\sim 1$ and smoothly and automatically asymptote to hydrodynamical evolution. 
}
\end{figure}

According to the picture arising from the saturation framework and CGC, the 
post-collisional debris that are left in the midrapidity region, and that will
eventually form the hydrodynamically expanding medium, consist of parametrically
overoccupied gluon fields at a typical saturation scale $Q_s$ \cite{Gelis:2010nm}.That is, the 
typical mode occupancies per unit phase phase space volume for gluons are 
inversely proportional to the coupling constant
\begin{equation}
\frac{d N_{gluons}}{d^3p \, d^3x }\equiv f \sim \frac{1}{\alpha_s}
\end{equation}
at the dominant momentum scale $Q_s$.
In contrast, a system that is in thermal equilibrium has phase space occupancies of the order of 
unity $f\sim 1$ at the scale given by the temperature $T$. The challenge to understand the process of thermalization at weak coupling is then
to understand how this overoccupied far-from-equilibrium state reaches the thermal state
under the violently expanding geometry of the heavy-ion collision.

The route the weakly coupled system takes to reach local thermal equilibrium is not straightforward. 
Indeed, it was first realized by Ref.~\cite{Baier:2000sb} that under longitudinal expansion a system that is initially overoccupied
becomes underoccupied before it reaches thermal equilibrium. This behaviour is depicted
in cartoon of Fig.~\ref{fig:route} on a plane of occupancy and anisotropy. The interactions
work to bring the system directly from the overoccupied initial condition to the thermal
equilibrium at the origin of the figure. Indeed, in the absence of expansion, the thermalization would
proceed along a straightforward route with monotonically decreasing occupancies directly to thermal equilibrium \cite{Kurkela:2011ti, Kurkela:2012hp,Schlichting:2012es,Kurkela:2014tea}.
The effect of the expansion, however, makes the system
more anisotropic. The combined effect of the two leads the system to miss the thermal 
equilibrium at the origin, and causes the system to flow to the underoccupied half of the diagram. This behaviour
is examined in detail in Refs.~\cite{Baier:2000sb,Kurkela:2011ti,Kurkela:2011ub}. 
Once the system turns underoccupied, its physics changes qualitatively (in particular for inelastic scattering which
changes from merging to splitting), and this change eventually allows the system to reach
the thermal equilibrium.

That the system explores both the underoccupied and overoccupied regions of the diagram
before settling to thermal equilibrium has a technical implication that the degrees of freedom
that are used to describe the system also change during the the prethermal evolution. 
On one hand, as long as the system is overoccupied, it can be described in terms of classical Yang-Mills (CYM) theory.%
\footnote{As long as the system is parametrically overoccupied, the fermions play only
a subleading role in thermalization as their occupancies cannot exceed 1 and therefore are always less numerous
as the gluons. } The corrections to classical field theory description comes from non-zero coupling
 $\alpha_s$ and from small occupancies $\mathcal{O}(f^{-2})$ \cite{Mueller:2002gd,Jeon:2004dh}. 
When the occupancies become less large, the ``quantum corrections'' arising from the expansion
in the inverse occupancy become large, and the when $f \sim 1$, the classical theory fails to give a faithful description of the system.  
Therefore the classical description can never reach the thermal equilibrium with $f \sim 1$. 

On the other hand, once the system becomes underoccupied, it is better described in terms of 
particle degrees of freedoms and their kinetic theory. The expansion parameters of the kinetic theory 
description are again the coupling constant itself, $\alpha_s$, but also the occupancies $\alpha_s f$.
Therefore the effective kinetic theory can be used to describe systems that are underoccupied, and the 
description can be extended to occupancies that are large but not non-perturbatively large $f \ll 1/\alpha_s$, so that it can reach the thermal equilibrium but not the CGC initial condition. 
\begin{figure}
\hfill
\includegraphics[width=0.4\textwidth]{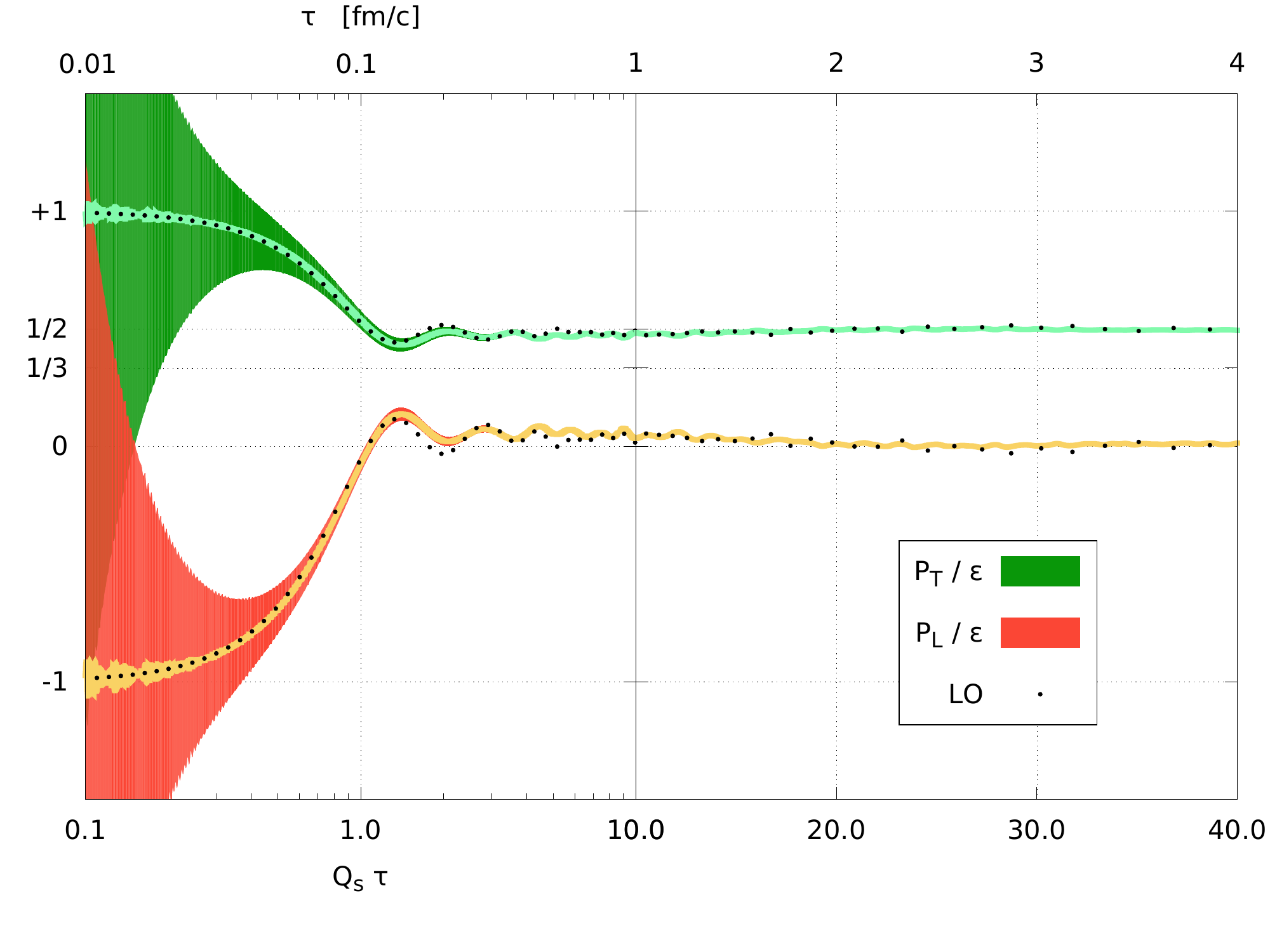}
\hfill
\includegraphics[width=0.35\textwidth]{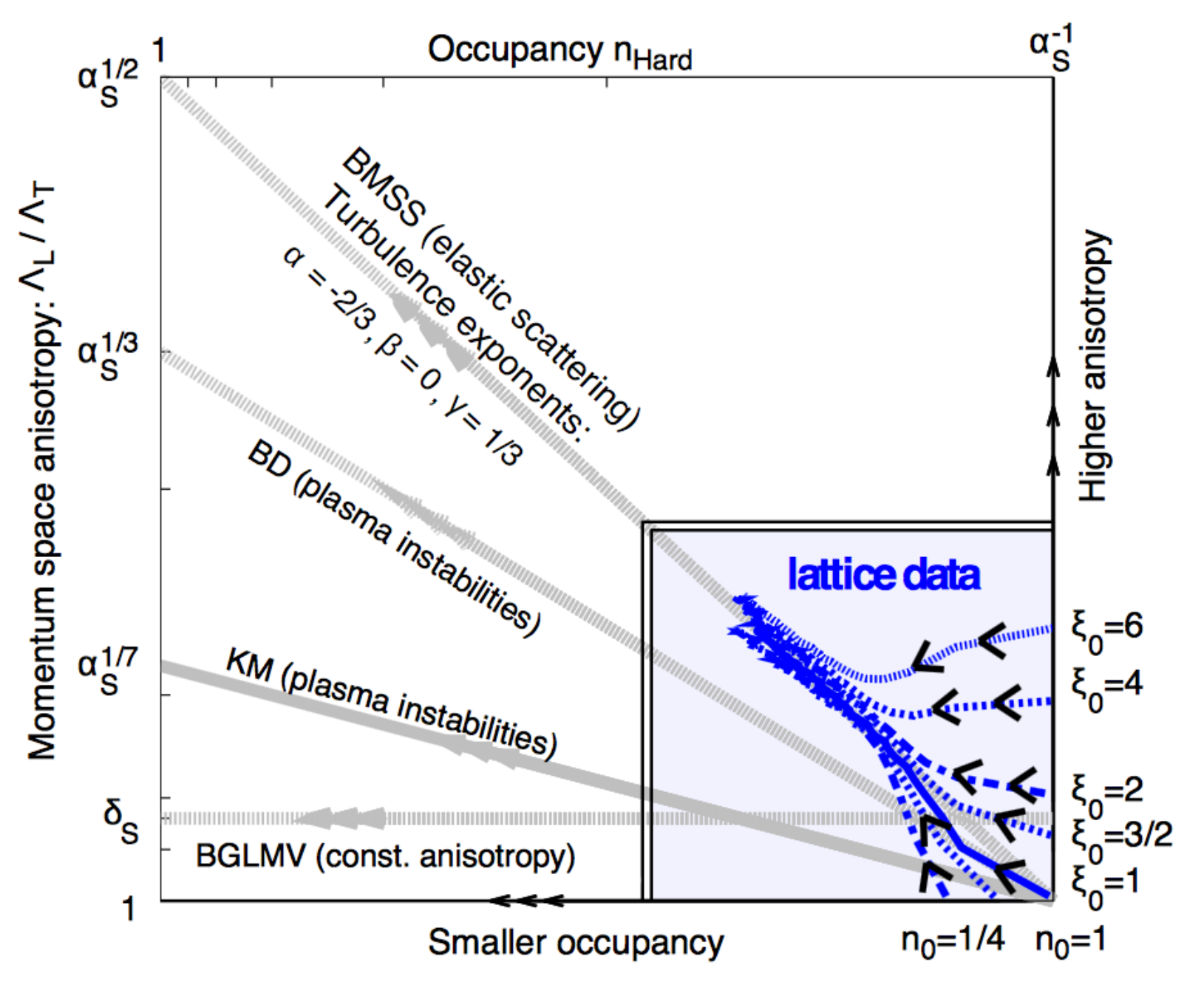}
\hfill
\vspace{-0.4cm}
\caption{
\label{fig:clas}
Classical evolution of the system. (Left) The time evolution of the components of energy momentum tensor from
a simulation initialized with CGC initial condition from \cite{Gelis:2013rba}. At late times the system becomes ever more anisotropic (Right) Trajectories of classical simulations in the plane of anisotropy and occupancy with different initial conditions \cite{Berges:2013fga}. All the trajectories become eventually more anisotropic and dilute irrespective of the details of the initial condition. The classical Yang-Mills theory never thermalizes or isotropizes.
}
\end{figure}

Neither of the descriptions, the classical field theory or the kinetic theory, can cover the whole 
time evolution of the system from the overoccupied initial condition to thermal equilibrium.  However,
a strategy where the early evolution is described by classical Yang-Mills theory and where 
the late approach to hydrodynamics is dealt within kinetic theory, can indeed bring the overoccupied initial
condition to thermal equilibrium. In such strategy, the system must be passed from
one set of degrees of freedom (the classical YM fields) to another (distribution functions of kinetic theory).
This transfer is made possible by the parametrically large region of overlapping validity
of both description. Indeed for
\begin{equation}
1 \gg f \gg 1/\alpha_s,
\end{equation}
both the CYM and the kinetic theory give a leading order accurate description of the non-equibrium
system. This region is denoted by the green area labelled ``Both'' in Fig.~\ref{fig:route}, where 
Fourier transforms of the classical fields can be interpreted as particle distributions.
Within this region, first evolving the gauge fields
according to the classical equations of motions and then Fourier transforming, is equivalent to first taking the Fourier transform of the fields and then evolving the distribution function in kinetic theory up to corrections that are $\mathcal{O}(f^{-2})$, and $\mathcal{O}(\alpha_s f)$.

A cartoon of the weak coupling strategy is displayed in Fig.~\ref{fig2}. At early times when the occupancies are non-perturbative $f\sim  1/\alpha_s$ the 
system is described in classical Yang-Mills theory. The classical evolution will quickly dilute the system and render the occupancies perturbative  $f \ll 1/\alpha_s$ . At any time $\tau_{EKT}$ during the window when the typical occupancies are large but perturbative $1 \gg f \gg 1/\alpha_s$, the system may be passed to the kinetic theory description. As both descriptions are equivalent in this region, any residual dependence on $\tau_{EKT}$ is a higher order effect.  The subsequent evolution in EKT will then deviate from the classical field evolution once the occupancies become close to unity $f\sim 1$ and eventually it will smoothly and automatically asymptote to hydrodynamical evolution. At this point, the system may be passed at any arbitrary time $\tau_i$ to a hydrodynamical 
description.

There has been significant quantitative progress in understanding the different pieces of this jigsaw puzzle, both
on the classical Yang-Mills side as well as on the kinetic theory side. In the rest of this contribution I will review
some of this progress and point out how the different works relate to the strategy outlined above.

\section{Classical field evolution}
\label{sec:clas}

The classical Yang-Mills simulations starting with various incarnations of CGC initial conditions
have appeared in the literature for a quite some time \cite{Krasnitz:1998ns,Krasnitz:1999wc,  Lappi:2003bi}.
More recent simulations include, e.g., JIMWLK renormalization group evolution \cite{Lappi:2011ju} of the initial condition. 
The CGC initial condition is boost invariant, and therefore the early simulations were performed in 2+1D space-time dimensions. 
 The boost invariance is, however, broken by fast growing (Chromo-Weibel) unstable modes seeded by 
 small fluctuations (with quantum or other origin), and modern simulations have included full 3+1D dynamics
\cite{Berges:2013eia,Berges:2013fga,Gelis:2013rba}.

Figure \ref{fig:clas} (left) from Ref.~\cite{Gelis:2013rba} shows the evolution of the transverse and longitudinal pressure starting from CGC initial condition
as described by the McLerran-Venugopalan model \cite{McLerran:1993ka}. The initial condition consists of coherent fields, leading
to negative longitudinal pressure. The coherence is quickly lost in a time scale of $1/Q_s$ and the longitudinal 
pressure becomes positive. At later times, the anisotropy $P_T/P_L$ grows consistent with the expectation
of Fig.~\ref{fig:route}.
%
The very early time evolution $\tau_{EKT}\lesssim 1/Q_s$ has been also studied analytically
in a recent paper \cite{Chen:2015wia}.

\begin{figure}
\begin{center}
\includegraphics[width=0.4\textwidth]{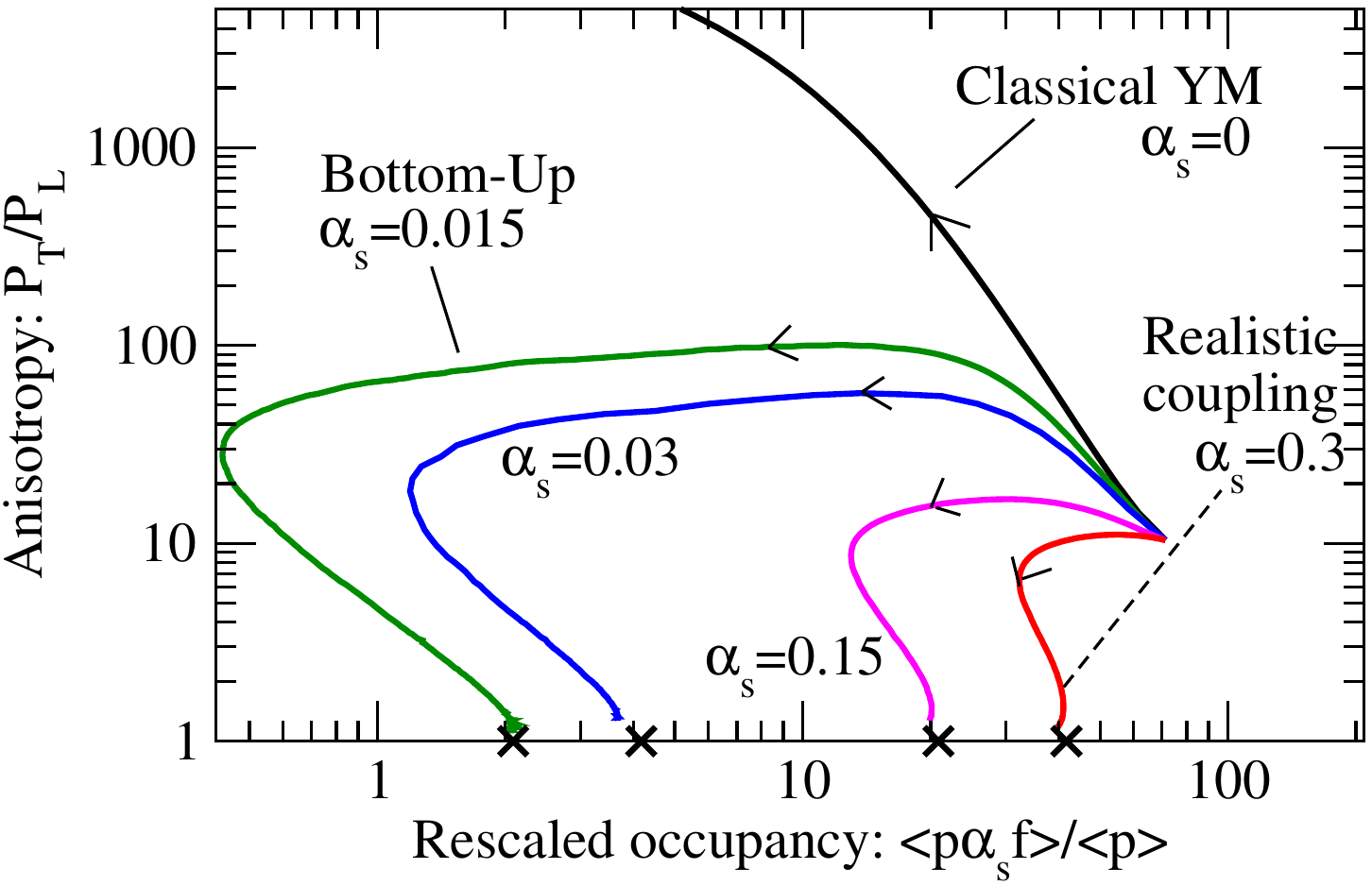}
\end{center}
\caption{
\label{fig:nroute}
Trajectories of effective kinetic theory simulations at different values of the coupling constant. The simulations start with an initial condition that is given by classical Yang-Mills simulation. The non-trivial limit of $\alpha_s \rightarrow 0$ with $\alpha_s f$ fixed corresponds to classical Yang-Mills limit and the trajectory becomes ever more anisotropic and dilute. At finite $\alpha_s$ the evolution differs from the CYM limit once $f\sim 1$ is reached, and the subsequent evolution brings the trajectory to thermal equilibrium marked with a black cross. 
}
\end{figure}

A complementary study \cite{Berges:2013eia,Berges:2013fga} discusses the late time limit of the classical evolution. It demonstrates the robustness
of the conclusion that the classical evolution leads to a system that becomes ever more dilute and 
anisotropic. The authors start their simulation with a broad range of different initial conditions, that is, with different
initial occupancies and anisotropies. It is observed that irrespective of the initial condition, the classical 
evolution leads to universal behaviour with occupancies decreasing and anisotropy increasing as a power law of proper time. This behaviour is depicted in Fig.~\ref{fig:clas} (right), where the different simulations
fall on a single trajectory on a plane of anisotropy and occupancy. This figure can be taken of as a numerical realization
of the right hand side of the Fig.~\ref{fig:route}. 

The classical dynamics will make the system ever more anistropic and dilute. At some point, however, the 
occupancies become of order unity and the classical approximation fails. This breaking is not, however, 
signalled in any way in the classical simulation; $\hbar$ does not appear anywhere in the classical 
simulation and the fields amplitudes can decrease ad infinitum. For a faithful description, however, before this failure one 
should switch to the kinetic theory description, whose behaviour changes qualitatively around $f\sim 1$.

\section{Classical particle evolution}
\label{sec:kin}

The effective kinetic theory (EKT) of quarks and gluons that is accurate to leading order in the coupling 
constant, is formulated in \cite{Arnold:2002zm}.
 The EKT is applied in a wide variety of dynamical non-equilibrium and transport phenomena, e.g., in the computation of leading order transport coefficients in QCD \cite{Arnold:2003zc} as well as jet phenomenology \cite{Schenke:2009gb}.  
The quantitative equivalence of the EKT with CYM in the overoccupied far-from-equilibrium setup
has been numerically demonstrated in an isotropic setup in \cite{York:2014wja}. 

Fig.~\ref{fig:nroute} displays the non-equilibrium evolution of a system within kinetic theory from \cite{Kurkela:2015qoa}. 
The initial condition for the kinetic theory evolution is a parametrization of a CYM simulation at a switching time $\tau_{EKT} = 1/Q_s$. The 
parametrisation is chosen to match energy density $\varepsilon$ and $\langle p_T \rangle$ with the classical simulation of \cite{Lappi:2011ju} at $\tau_{EKT}$. The figure contains 
trajectories from simulations with different values of the coupling constant $\alpha_s$. The x-axis is rescaled with $\alpha_s$ such that  
for all values of $\alpha_s$, overoccupied initial condition with $f\sim 1/\alpha_s$ falls on the same point in the plane of occupancy and 
aniostropy. With such a rescaling, taking the limit $\alpha_s \rightarrow 0$ keeping the initial $\alpha_s f$ fixed, gives a non-trivial limit that 
corresponds to the CYM. In terms of the figure, this limit corresponds to sending $f\sim 1$ to infinitely far to the left, so that occupancies of order one are never reached in a simulation. This trajectory is labelled ``classical YM'', and indeed what is seen is consistent with CYM simulations. The system 
becomes ever more dilute and anisotropic and never thermalizes.

For a small but non-zero $\alpha_s$, the kinetic theory simulation eventually reaches $f\sim 1$, and the qualitative change in the dynamics
is clearly visible in the line labelled ``Bottom-Up'' in Fig.~\ref{fig:nroute}. When the occupancies become small, the EKT simulation deviates from the classical limit and the growth of the anisotropy stops. Eventually the EKT evolution brings the trajectory close to thermal equilibrium  Fig.~\ref{fig:nroute} as anticipated by \cite{Baier:2000sb} and consistent with expectation of Fig.~\ref{fig:route}.

Increasing the coupling further does not lead into further qualitative change. At larger couplings, the features in the trajectory arising from 
the scale separations provided by the weak coupling do become less pronounced and the system takes a somewhat more straightforward path to equilibrium. 

\begin{figure}
\begin{center}
\includegraphics[width=0.45\textwidth]{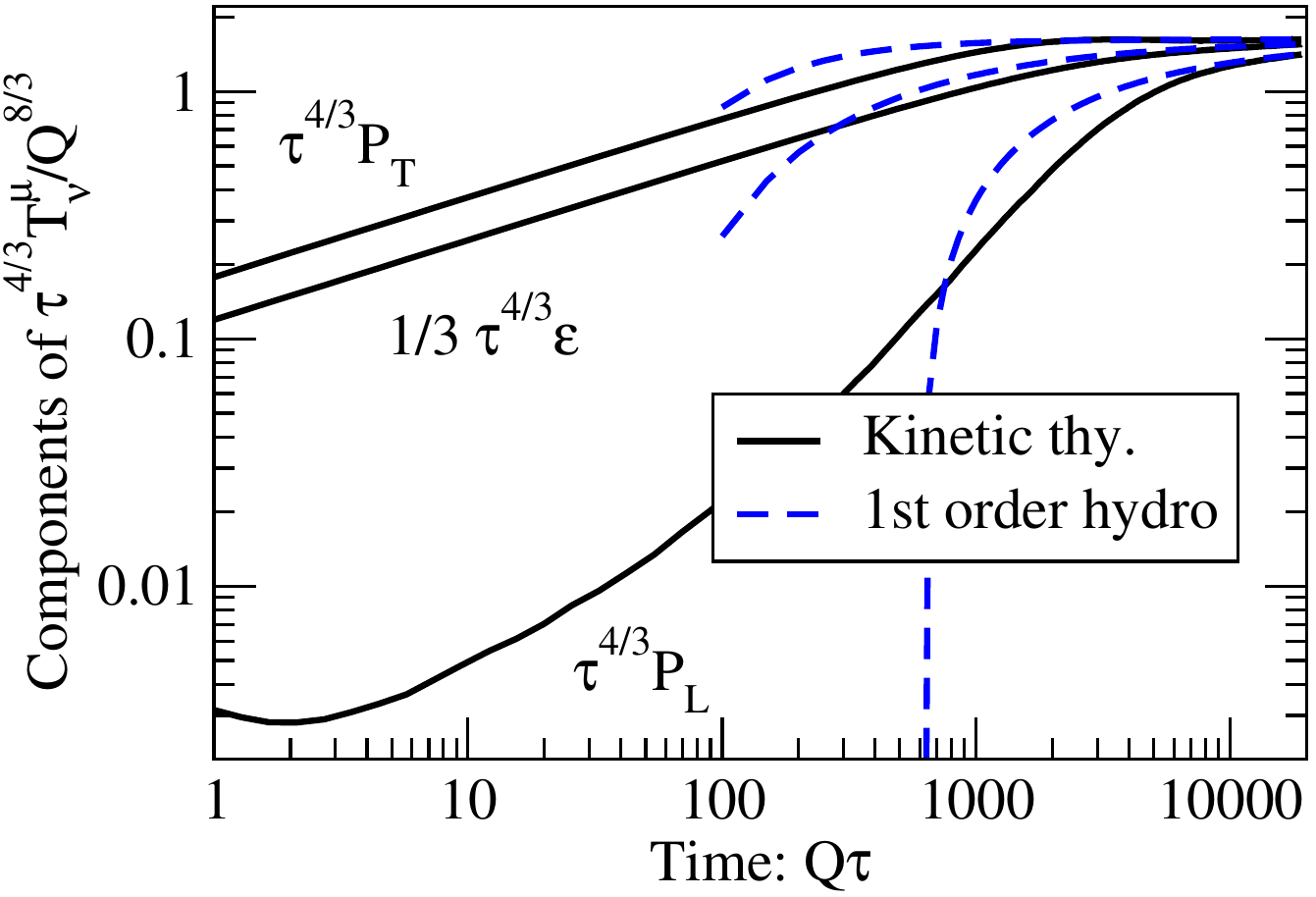}
\hfill
\includegraphics[width=0.45\textwidth]{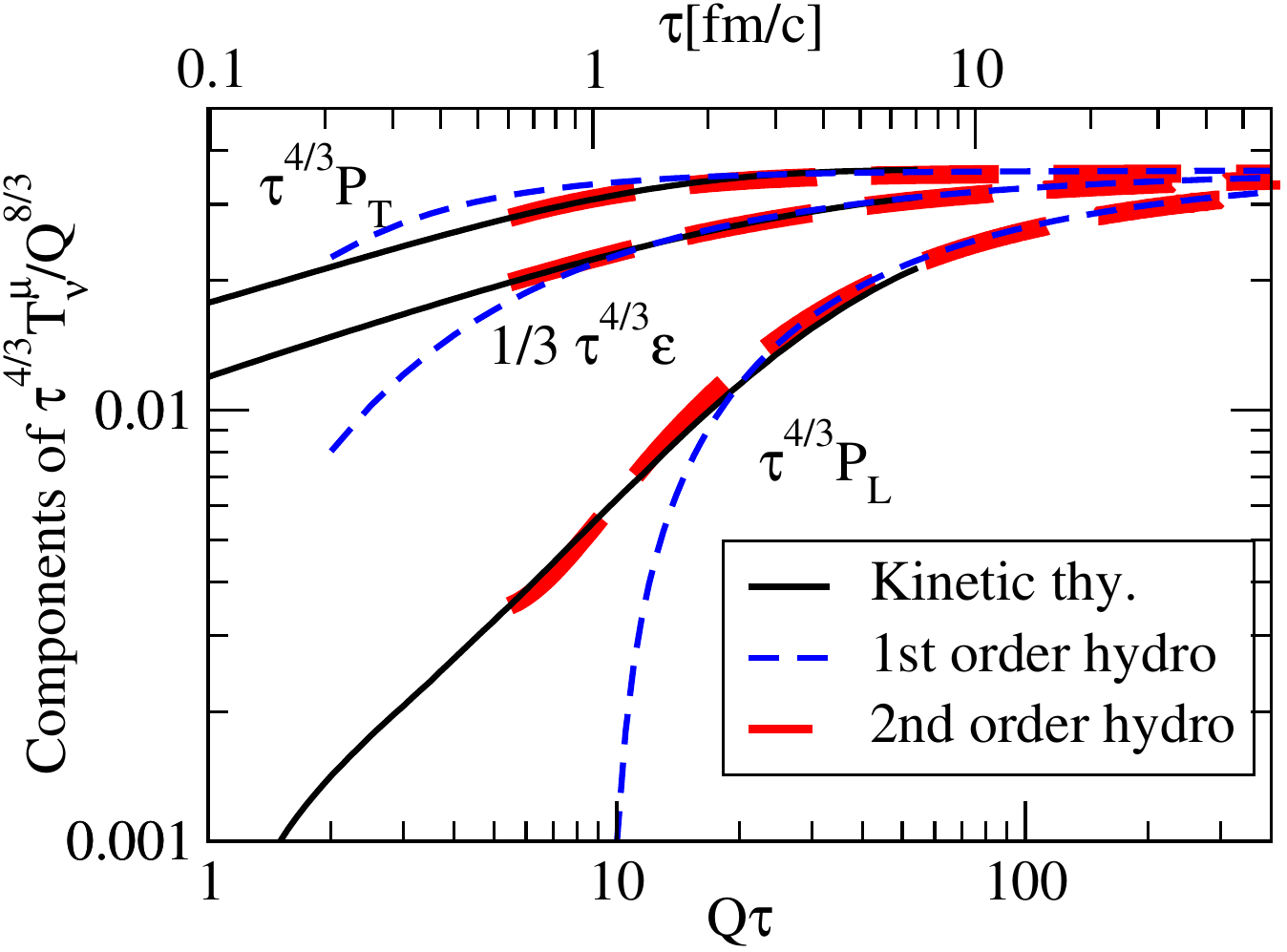}
\end{center}
\vspace{-0.4cm}
\caption{
The time evolution of the components of the energy-momentum tensor from EKT simulations with $\alpha_s \approx 0.03$ (left) and $\alpha_s \approx 0.3$ (right). The y-axis is rescaled by $\tau^{4/3}$ so that the horizontal lines correspond to ideal hydrodynamics. The approach to hydrodynamics is governed by parameter free prediction of viscous hydrodynamics (blue dashed lines). 
\label{fig:hydro}
}
\end{figure}

\subsection{Hydrodynamization}
As the kinetic theory evolution brings the system close to thermal equilibrium, it should at late times be described by hydrodynamics. 
This is illustrated in Fig.~\ref{fig:hydro}, which depicts the time evolution of the components of the energy momentum tensor from 
a simulation with $\alpha_s \approx 0.03$ (left) and $\alpha_s \approx 0.3$ (right). The energy
density $\varepsilon$ and the transverse and longitudinal pressures $P_T$ and $P_L$ are normalized by $\tau^{4/3}$. With this normalization, if the evolution is described by ideal hydrodynamics (with conformal equation of state ) for which $\varepsilon/3= P_T = P_L \propto \tau^{-4/3}$, the lines should be horizontal and fall on top of each other. It can be clearly seen from the figure that indeed at late times, the 
lines asymptote to horizontal lines, and at late times the system is indeed described by ideal hydrodynamics.

According to the hydrodynamical theory, the approach to ideal hydrodynamics should be described by viscous hydrodynamics, which gives a prediction that depends only on
the shear viscosity $\eta$ of the matter. As this is a well known quantity at weak coupling \cite{Arnold:2003zc}, the comparison with 
the parameter free prediction of hydrodynamics provides a very robust test. 
The prediction
from viscous hydrodynamics is shown as the blue dashed line in Fig.~\ref{fig:hydro}. The approach is
well captured by the hydrodynamical prediction for both couplings shown. For the small coupling $\alpha_s \approx 0.03$,
the viscous hydrodynamics agrees with the kinetic theory simulation at very late times $\tau \sim 5000/Q_s$. The 
simulation with intermediate coupling $\alpha_s \approx 0.3$ shows qualitatively similar features but differs quantitatively; it follows the hydrodynamical line already around $\tau \sim 10/Q_s$.
Converting this to physical units using a typical estimate of $Q_s \approx 2$GeV gives hydrodynamization times of $\sim 1$fm/c. 
It is quite remarkable that the hydrodynamical theory works already at this time, when the anisotropies are still quite large $P_L/P_T \sim 0.2$.

\section{Discussions}
In these proceedings I have discussed the not-so-straight-forward way the heavy-ion collisions thermalize, isotropize,
and reach hydrodynamical flow in the limit of weak coupling. The system undergoes two smoothly connected stages before hydrodynamizing, the overoccupied and the underoccupied stages. These stages must be described in terms of different degrees of freedom, either in terms of classical fields or in terms of (semi-)classical particles. 
The thermalization times, obtained from the weak coupling framework extrapolated to intermediate couplings show agreement with the phenomenological expectation of $\sim$1fm/c.

The effective kinetic theory framework that describes the thermalization is used also to describe physics at widely different energy scales in the collision such as photon production and jet quenching. As such, the unifying kinetic theory prescription and the early time dynamics builds a bridge between the soft physics of flow and hard probes. 

At present there are still several caveats that need to be ironed out before the model is ready for realistic phenomenology. So far the 
studies do not include fermions, which play a subleading role in the process of thermalization due to fermi statistics and color factors. 
However, the thermalization of fermions is crucial, for example, to electromagnetic probes, and if the chemical equilibration happens delayed with respect to kinematic thermalization, one could question whether the 2+1 flavour equation of state is the right EoS to use at early times in hydrodynamical simulations. The current implementation of the EKT does not allow for transverse dynamics, which is a crucial for the preflow, and the numerical role of certain plasma unstable modes is still somewhat unclear. Also, further work in the interface between the CYM and EKT is needed. However, while there are still several caveats, the list is getting shorter and the field of weak coupling thermalization is becoming a quantitative field. 

How much of this theoretical exercise performed at the weak coupling limit carries through to physical heavy-ion collisions that are performed at RHIC or at the LHC, which unfortunately do not run at the asymptotic limit $\sqrt{s}\rightarrow \infty$? Are we allowed to extrapolate the weak coupling results to the relevant interesting intermediate couplings around $\alpha_s = 0.3$? A pessimist view would be to note that perturbation theory at finite temperature is notoriously badly convergent at temperatures close to $T_c$ \cite{Kajantie:2002wa}. However, an optimist would notice that as the thermalization takes place rather fast, the resulting temperatures at $\tau_{EKT}$ are still of order $3-4T_c$. This does lie in the temperature range where the perturbation theory in its modern incarnations seems to describe lattice results. The hydrodynamical evolution will eventually cool the system to a temperature range where a weak coupling description of the medium most likely is inappropriate, but by then the system has been already passed to a hydrodynamical description. As the hydrodynamical analysis mostly 
constrain the minimum of $\eta/s$ around $T_c$ \cite{Niemi:2015qia}, not its value around $3T_c$, this picture is not in contrast with the phenomenology and the paradigm of small $\eta/s$. 

So far the strategy has been to extrapolate the weak coupling results to intermediate couplings expected to realize in physical collisions. The same goes for the strong coupling studies where the results obtained at infinite coupling are again extrapolated down to intermediate couplings. However, perhaps a better strategy than extrapolating would be to use all available information to constrain the intermediate couplings; to interpolate rather than extrapolate. First steps in this direction are taken in \cite{Keegan:2015avk}, where non-equilibrium evolutions of strongly and weakly coupled theories with similar initial conditions are compared.  

Whether the strategy is to extrapolate or to interpolate in coupling, it is clear that it is of importance to push the calculations beyond the 
leading order accuracy, both at weak and at strong coupling in order to increase precision of the results and to better evaluate at what range of couplings they provide useful constraints. While it
will be a long way to NLO, there have been important advances in the recent past to this direction \cite{Ghiglieri:2015ala}.












\end{document}